\documentclass{article}
\usepackage[utf8]{inputenc} 
\usepackage[backend=biber, style=authoryear]{biblatex}
\usepackage{xcolor,colortbl}
\usepackage{PRIMEarxiv}
\usepackage{geometry}
\usepackage{wrapfig}
\usepackage{subcaption}
\usepackage[T1]{fontenc}    
\usepackage{hyperref}       
\usepackage{url}            
\usepackage{booktabs}       
\usepackage{amsfonts}       
\usepackage{nicefrac}       
\usepackage{microtype}      
\usepackage{lipsum}
\usepackage{fancyhdr}       
\usepackage{graphicx}       
\graphicspath{{media/}}     
\usepackage{tabularx}
\usepackage{float}
\usepackage{amsmath}
\title{Analysis of points outcome in ATP Grand Slam Tennis using big data and machine learning
}

\author{ 1. Martin Illum \\
	Technical University of Denmark\\
	\texttt{s184286@student.dtu.dk} \\
	\And
    2. Hans Christian Bechsøfft Mikkelsen \\
	Technical University of Denmark\\
	\texttt{s184294@student.dtu.dk} \\
	\And
	3. Emil Hovad \\ 
Technical University of Denmark \\
	\texttt{emilh@dtu.dk} \\
}

\begin{document}
\maketitle

\begin{abstract}
Tennis is one of the world's biggest and most popular sports. Multiple researchers have, with limited success, modeled the outcome of matches using probability modelling or machine learning approaches. The approach presented here predicts the outcomes of points in tennis matches. This is based on given a probability of winning a point, based on the prior history of matches, the current match, the player rankings and if the points are started with a first or second. The use of historical public data from the matches and the players' ranking has made this study possible. In addition, we interpret the models in order to reveal important strategic factors for winning points. The historical data are from the years 2016 to 2020 in the two Grand Slam tournaments, Wimbledon and US Open, resulting in a total of 709 matches. Different machine learning methods are applied for this work such as, e.g. logistic regression, Random forest, ADABoost, and XGBoost. These models are compared to a baseline model, namely a traditional statistics measure, in this case the average. An evaluation of the results showed that the models for points proved to be a fraction better than the average. However, with the applied public data and the information level of the data, the approach presented here is not optimal for predicting who wins when the opponents are on the same position on the ranking. This methodology is interesting with respect to examining which factors are important for the outcomes of who wins points in tennis matches. Other higher quality data sets exists from e.g. Hawk Eye, although these data sets are not available for the public.
\end{abstract}


\section{Introduction}
Tennis is one of the world's most popular sports, with an estimate of a billion fans worldwide \cite{2021TopWorld}. The Association of Tennis Professionals (ATP) Tour features more than 60 tournaments in over 30 countries with over 1500 professional players each year, which makes the competition really tough. Many players are therefore striving to gain an edge over their opponents. This causes an ever increasing demand for scientific research, mainly through data science, on how players can achieve success.

Baseball was one of the first sports where large quantities of data were collected thereby  statistics and data science could be applied, also called the study of Sabermetrics. Bill James heavily introduced Sabermetrics during the 1970 to 1980's \cite{James2003}. Continuing with the baseball team Oakland Athletics consistently appeared in the Major League Baseball playoffs in the early 2000's even though, they were amongst the teams with the lowest salary payroll in the entire league. The Oakland Athletics achieved these results, by redefining how they evaluated and scouted players, as discussed in the book \cite{Lewis2003}. 

 Rafeal Nadal the 20 times Grand Slam winner said just before the Wimbledon final in 2008: \textit{"If I have to hit a ball twenty times to Federer’s backhand, I’ll hit it twenty times, not nineteen. If I have to wait for the rally to stretch to ten shots or twelve or fifteen to bide my chance to hit a winner, I’ll wait. There are moments when you have a chance to go for winning drive, but you have 70 percent chance of succeeding; you wait five shots more and your odds will have improved to 85 percent. So be alert, be patience, don’t be rash."} \cite{TennisNadal}. Continuing in this track former Team Djokovic strategy analyst Craig O'Shannessy thinks that the term "\textit{Moneyball}" can be connect with modern tennis \cite{OShannessy2019TennisTennis}. Although, non-public available private vendor data exists with fine-grained spatiotemporal data from professional tennis matches at "\textit{www.atptour.com}". Here it is possible to see analysis of tennis matches, but to our knowledge it is not possible to access or download the data for additional investigation or research. Altough, analysis was perform with respect to serve return patterns \cite{Kovalchik2022} based on the "\textit{www.atptour.com}" dataset, in this study there are no links to the data. The same sport scientist Stephaine Kovalchik thinks that tennis is lacking behind other sports, like baseball and due to the lack of data describing the actual gameplay, it is to early to say that tennis is in the "\textit{Moneyball}" era \cite{Kovalchik2021WhyMoneyball}. 
 
Betting sites does sometimes provide public data as "\textit{http://tennis-data.co.uk/data.php}" and this data has been used for research to find the greatest player ever  \cite{Kovalchik+2016+127+138}. Data for tennis can sometimes be found the Kaggle homepage "\textit{https://www.kaggle.com/code/residentmario/exploring-world-tennis-matches}" with the title "Exploring World Tennis Matches". This dataset mostly provide information about when and where tennis matches have been played. Sometimes detailed data does exists, altough only for one match, in this case Tennis ATP Tour Australian Open Final 2019 at "\textit{https://www.kaggle.com/datasets/robseidl/tennis-atp-tour-australian-open-final-2019}", where events was collected as e.g. when a player hit the ball, the stroke type, position of the player, and position of the opponent were recorded.
This would be somewhat possible if The Hawk Eye data was merged with Jeff Sackmann's data. 

The availability of public tennis data has been limited, this data set from Jeff Sackann \cite{JeffSackannPbP, JeffSackannATP} is currently to our knowledge the most used public dataset and is a collection of data from several sources.

This data is used to investigate if data from ATP Grand Slam tennis matches can be used to predict the outcome of tennis points, with various machine learning methods. This approach will lead to the following problem statement: can we find important factors also called features for winning in tennis and predict who wins the points. There is a significant difference between the male and female tennis beside 3 sets vs. 5 sets, this can also be seen in the serve, which is a more deciding factor for male tennis than for female tennis matches \cite{Carboch2017Comparison2016}. Therefore, we only focus on male matches in this paper.



For the future a potential tennis "\textit{Moneyball}"  dataset would be a dataset which collected information about the points outcomes, timestamps, type of stroke, position of the ball and the player in 3-D. Additionally if static videos (video camera is static) from different angles of the matches could be available as well potentially with tracking information in pixel level and potentially with triangulation as well together with calibrated cameras.

\section{Methods}\label{Methods}
In this section the creation of the point Winner model is outlined for the first serve and second serve. Two different approaches were used to create these two similar models. The point winner model, we predict who wins the next point based on the serve being a first or second serve, from any given point in the match taking into account previously played points. 

Furthermore, we use a baseline model and we make a comparison to four machine learning methods for the point winner model. These machine learning methods are chosen based on their different levels of complexity, and their ability to explain the data vs. prediction ability. The hyperparameters are optimized for the best performing models in the validation data, by random search.

\subsection{Baseline model}
The baseline models are calculated by the number of times the player wins the serve divided by the total amount of serves for the first serve or for the second serve. This is calculated independently for the first and second serve, these are seen as different events and thereby two models are made for these two scenarios.

\subsection{Logistic regression Classifier}
Logistic Regression Classifier is a supervised machine learning technique, that can be used to prediction of binary classes this gives probabilities for the outcomes. In this case if the server wins or loses the point. Logistic regression is an extended version of linear regression. 
\begin{equation}
    G({\mathbf{x}}) =  \frac{1}{ 1 + e^{-\mathbf{w}^{t}\mathbf{x}}} 
\end{equation}
A transformation function $G(\mathbf{x})$ is used on the input $\mathbf{x}$, called the Sigmoid function which obtains a probability by transforming the output of traditional linear regression to a value between 0 and 1. The parameter learnt is $-w^t$. The loss function, $L(G(\mathbf{x}),y)$ is function is defined by
\begin{equation}
    L(\mathbf{x}) =  \frac{1}{m}  \sum_{i=1}^{m}   - (1-y^{i})log(1+y^{i}-G(\mathbf{x}^{i})) - y^{i} log(G(\mathbf{x}^{i}))
\end{equation}
Where $y$ is the target value and $m$ is the number of training examples. The update of $\mathbf{w}^{t}$ is typically performed by stochastic gradient descent, decreasing the loss with a step the learning rate $\alpha$.
\begin{equation}
    w_{j,n+1} =  w_{j,n} - \alpha \frac{\partial  }{\partial w_{j}}L(\mathbf{x})
\end{equation}
Some of the first application of the Logistic regression is in e.g.
\cite{10.2307/2280041}.



\subsection{Random forest Classifier}
A single tree also called a decision tree is created by making decision boundaries for dimension of the data set to make partitions for the different classes. You find the most important dimension of the data (the best questions or the first leave), which gives the best split of the data with the lowest entropy by the impurity function.   

Random Forest is an ensemble tree-based learning algorithm. It is an algorithm which consist of a collection of tree classifiers grown using bagging ${M(\textbf{x},\Theta_k), \textit{k}=1,...}$, where $\Theta_k$ are independent identically distributed random vectors, where the final result is found by majority voting of all tree classifiers. Compared to regular decision trees, random forest adds an additional element of randomness, while growing the trees. When trees are split, it finds the most important feature in a random subset of features instead of all features, which results in a more diverse model, compared to a normal decision tree, which searches for the most important feature of all the features \cite{Breiman2001RandomForests}. 

A few advantages of using random forest in this paper is, that it handle both categorical and continuous values well, and the random element in the algorithm contributes to less overfitting. The model is however more computational heavy, and its suffers from bad interpret-ability, by using weights for the features, it is not possible to determine whether a feature has a negative or positive impact for the result.

\subsection{AdaBoost Classifier}
Adaboost is a boosting method, which combines weighted weak learners (classifiers) and assigning them weights depending on how precise they classified the observations as compared to the equally weighted Random forest classifier. The weak learners are made according to how well the previous weak learner classified the observations \cite{FREUND1995256} and the badly predicted observations are weighted higher when the next classifier is created and trained, after training all the weighted weak classifier are ready for the final classifier. For a more detailed description of the algorithm of AdaBoost classifier, see \cite{hastie01statisticallearning}. 

\subsection{XGBoost Classifier}
Extreme gradient boosting (XGBoost), builds upon the foundations of classical gradient boosting \cite{Chen2016XGBoost:System,hastie01statisticallearning}. It's a scalable machine learning algorithm for tree boosting, which is used both for classification and regression problems. It ensembles multiple weak classifiers, which are added sequentially and each focuses on correcting the mistakes from the prior classifier by minimizing a differentiable loss function. The objective is then to grow new trees which combined with the existing trees, lowers the total loss compared to the existing trees alone, while penalizing higher complexity. It is an additive approach which converts weak classifiers into a single strong classifier. The final prediction of the model is the sum of each tree.

\subsection{Evaluation Metrics}
In order to evaluate the performance of the applied machine learning models in this project, several evaluation metrics will be introduced. Different metrics are suitable for different type of problems and data structures, where the problem faced in this work is a binary classification problem.

\subsubsection{Confusion Matrix}
The confusion matrix is a $N \times N$ dimensional matrix, where $N$ is the number of classes to be predicted, which describes the performance of a model. From the confusion matrix multiple metrics can be derived, which are useful to analyze, how a model is performing. Here unbalanced data are encountered, therefore the metrics precision, recall, F-1 Score and Roc-Auc score will be added in addition to the accuracy score. These metrics are chosen, since they do well with unbalanced data sets, where traditional accuracy can be misleading, accuracy is however kept for an overall score of the model performances.

\subsection{Software Tools}
The programming language used for processing data and modelling is Python and Jupyter Notebook is used to present the code. Jupyter Notebook is an open-source software for Python and other programming languages. In Python, the main libraries used for the machine learning models and visualisations are Matplotlib, Numpy, Pandas, Scikit-learn and Seaborn.

\section{Data}
The data used for this study is available from the two public tennis data resources:
\cite{JeffSackannPbP} and \cite{JeffSackannATP}. The male matches are in a best-of-five set format and the female matches are in a best-of-three set format. 
We only use the male 5 sets matches and the \cite{JeffSackannPbP} data contains information for Grand Slams matches from 2011 to 2020. Only data from Wimbledon and US Open from 2016 to 2020 will be used, because these years and tournaments have additional information about the placements of the serves and returns.

\subsection{Data preparation}
A throughout data preparation has been carried out. The first step was to merge and sort out all the different data sets. Secondly was handling missing values in the matches, player ranks etc. Specially the features describing the placements of the serves was missing for some matches, therefore all these matches were excluded. 

\subsubsection{Accumulated features and restructuring of data}
Each row represents a point in the data table, what has occurred in it and what the score are afterwards. Multiple features such as \textit{P1UnfErr}, directly indicates which player wins the point and sometimes the game or set and therefore will these features not be used in the models. Instead accumulated features for each feature are created, which directly indicate which player wins the point. By including accumulated features information is also kept of how a player has performed up until the given point in a match. The new accumulated features and additional other features describing the score of a match etc.  \textit{P1Score}, \textit{P1GamesWon}, still shows the outcome at the end of a point and not the start of it. This problem is dealt with by shifting all column containing information regarding the outcome of a point one row up. The final result is that each row contains the information from a point up until the serve has been made, as well as how the players have performed during the whole match given by the accumulated features.

To make it easier to analyze and interpret the results of the models, the serving player will always be identified as player 1 and the returning player as player 2. The reasoning behind this is to avoid having two almost identical classes inform of the two players in a given match. Using this approach also makes it possible to see how serving and returning specific features affect the model. This is technically carried out by taking all player specific features such as \textit{P1PointsWon} and \textit{P2PointsWon} etc. and swapping the values between them, so when when player two is serving, he will appear as player one in the data set and vice versa. 

\subsection{Match patterns}
\begin{figure}[h]
\begin{subfigure}{.5\textwidth}
  \centering
\includegraphics[width=0.95\linewidth]{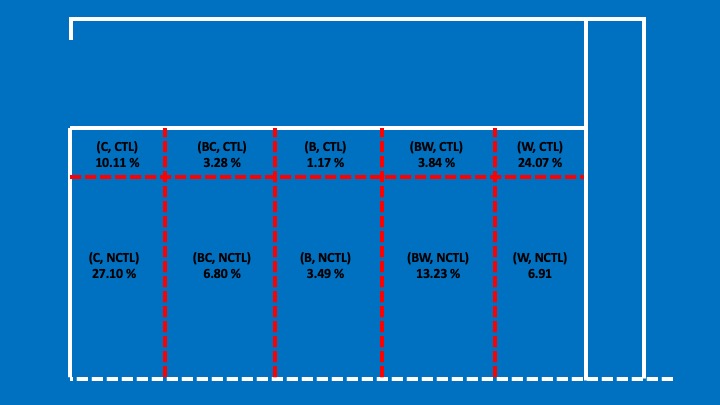}  
  \caption{1. Serve from AD court}
  \label{fig:1stServe1}
\end{subfigure}
\begin{subfigure}{.5\textwidth}
  \centering
  \includegraphics[width=0.95\linewidth]{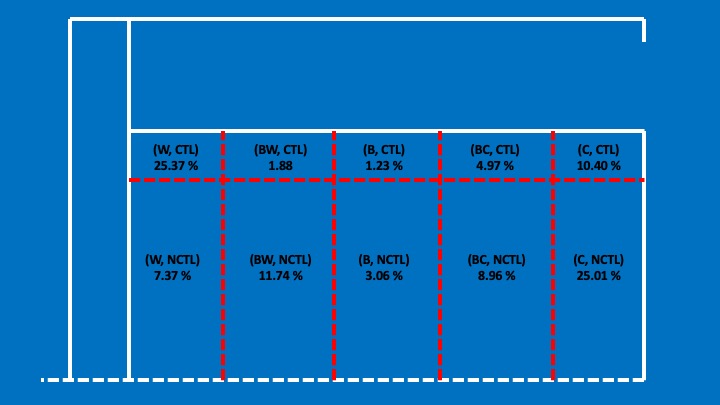}  
  \caption{1. Serve from Deuce court}
  \label{fig:2ndServe}
\end{subfigure}
\caption{Images of one-quarter of a tennis court with  combinations of ServeWidth and ServeDepth. Image (a) shows the combinations of ServeWidth and ServeDepth for the 1st serve from the AD court, and image (b) shows the combinations of ServeWidth and ServeDepth for the 1st serve from the deuce court}
\label{fig:DeltaFig1}
\end{figure}

\begin{figure}[h]
\begin{subfigure}{.5\textwidth}
  \centering
\includegraphics[width=0.95\linewidth]{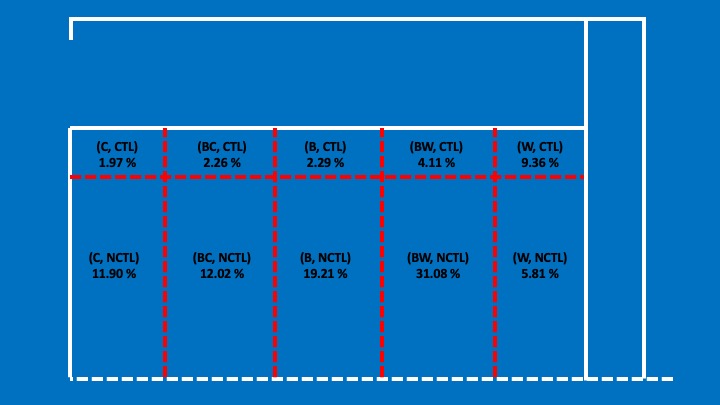}  
  \caption{2. Serve from AD court}
  \label{fig:1stServe}
\end{subfigure}
\begin{subfigure}{.5\textwidth}
  \centering
  \includegraphics[width=0.95\linewidth]{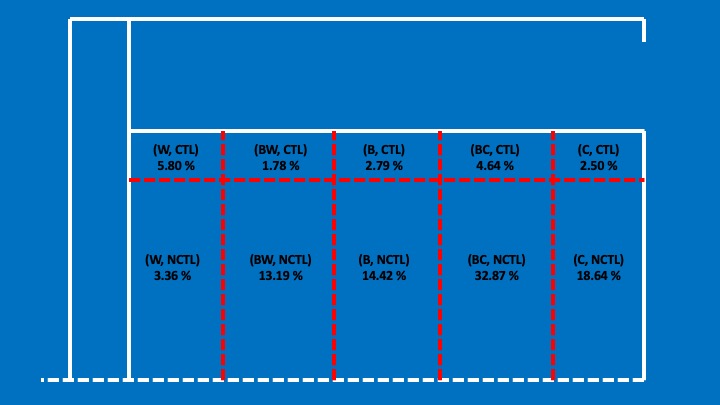}  
  \caption{2. Serve from Deuce court}
  \label{fig:2ndServe from}
\end{subfigure}
\caption{Images of one-quarter of a tennis court with  combinations of ServeWidth and ServeDepth. Image (a) shows the combinations of ServeWidth and ServeDepth for the 2nd serve from the AD court, and image (b) shows the combinations of ServeWidth and ServeDepth for the 2nd serve from the deuce court}
\label{fig:DeltaFig2}
\end{figure}

The importance of the serve in tennis is well known. So it would be interesting to see where players tend to serve for their first and second serve. In Figure \ref{fig:DeltaFig1} the frequency in percentage of all successfully first serve placements are shown for both the deuce and ad court. It can be seen that 68.19\% of the successful serves were served close to the sidelines of the service box for the AD court and 68.15\% for the deuce court. If compared to the second serve, seen in Figure \ref{fig:DeltaFig2}, where only 29.04\% of the serves were served close to the sidelines of the service box for the AD court and 30.3 from the deuce court. Furthermore it can be seen in Figure \ref{fig:DeltaFig2} that four of the most occurring serve positions had a serve depth of NCTL. This is most likely because if the second serve is unsuccessful the returner will get the point. Therefore the server is not willing to take risks when serving the second serve. It is important to mention that Figure \ref{fig:DeltaFig1} and \ref{fig:DeltaFig2} is an estimation of the placements for the combinations of the \textit{ServeWidth} and \textit{ServeDepth}, since the data does not specify the exact coordinates of the serve and return.

It can be seen that there was a difference between the placement of the serves from first to second. The second serves placement were often worse than the first serve. This affects the win percentage for the server. In the data the server wins 73.2\% of the points where the first serve is successful. On the other hand will the server only win 57.2\% of the points were the second serve is successful. A comparison to the overall server win percentage at 64.8\%, shows the importance of hitting the first serve.

\subsection{Procedure for Model Creation}
\begin{figure}[H]
    \centering
    \includegraphics[width=1\linewidth]{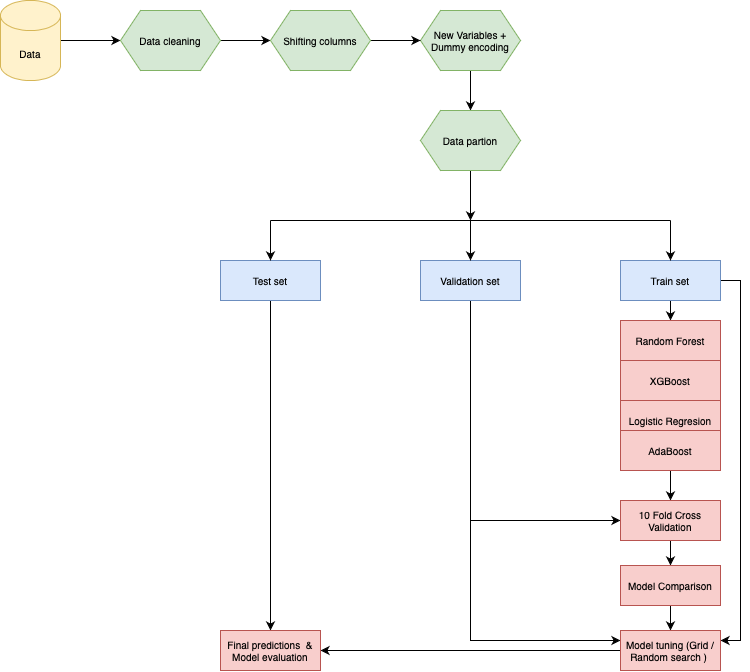}
    \caption{General approach for model creation}
    \label{fig:Flow}
\end{figure}

The general approach to the model creation can be seen in Figure \ref{fig:Flow}. First the data is collected, then the data preparation conducted by removing NaN-values, creating new features, one-hot encoding and data standardization. The data is then split into three subsets, where the test data consist of 10\% of all the matches randomly chosen, and is saved to test the final models. The validation and training data consist of the rest of the 90\% of the data, which is split by the ratio 80\% / 20\%. The four machine learning methods are then trained and validated upon these using 10-fold cross-validation. The performance of each model is then compared on evaluation metrics presented in the \nameref{Methods} section. The best overall performing model is then chosen and its hyperparameters are optimized using random search and is applied to the test data for a final evaluation.

\subsubsection{Model Features}
There are 57 features used as input for the models. These features includes accumulated features which feeds the model the historical data of a match, general features which e.g. describes the surface which the match is being played upon or the ranking of the players, and finally match specific features, e.g whether the point being predicted is part of a tiebreak. The features are further described in table \ref{FeatureD}

\begin{table}[h]
\centering
\caption{\label{FeatureD} Feature description of the data columns from \cite{JeffSackannPbP, JeffSackannATP}.}

\tiny
\footnotesize
\begin{tabularx}{\textwidth}{|l|l|X|}
\hline
\textbf{Feature}& \textbf{Type} & \textbf{Explanation}     \\ \hline
Match\_id & Discrete and nominal &  A unique number identifying a given match.\\ \hline
ElapsedTime & Discrete and interval & The start time of a point.    \\ \hline
SetNo & Discrete and ratio & The set that is being played for.      \\ \hline
P1GamesWon & Discrete and ratio & Number of games player 1 has won.   \\ \hline
P2GamesWon & Discrete and ratio & Number of games player 2 has won.     \\ \hline
SetWinner & Discrete and Ordinal & Which player won the set.     \\ \hline
GameNo & Discrete and ratio & The game that is being played for in the given set.    \\ \hline
GameWinner & Discrete and Ordinal & Which player won the game. \\ \hline
PointNumber & Discrete and ratio & What point is being played for.    \\ \hline
PointWinner & Discrete and Ordinal & Which player won the point.     \\ \hline
PointServer & Discrete and Ordinal & Which player serves.     \\ \hline
Speed\_KMH & Continuous and ratio & Speed of the serve in kilometers per hour.    \\ \hline
P1Score & Discrete and Ordinal & Player 1's score when the point start . \\ \hline
P2Score & Discrete and Ordinal & Player 2's score when the point start .    \\ \hline
P1PointsWon & Discrete and ratio & Number of points player 1 have won.   \\ \hline
P2PointsWon & Discrete and ratio & Number of points player 2 have won.   \\ \hline
P1Ace & Discrete and nominal & Indicator if player 1 has made an ace.  \\ \hline
P2Ace & Discrete and nominal & Indicator if player 2 has made an ace.  \\ \hline
P1Winner & Discrete and nominal & Indicator if player 1 has made a winner.  \\ \hline
P2Winner & Discrete and nominal & Indicator if player 2 has made a winner.  \\ \hline
P1DoubleFault & Discrete and nominal & Indicator if player 1 has made a double fault.  \\ \hline
P2DoubleFault & Discrete and nominal & Indicator if player 2 has made a double fault.  \\ \hline
P1UnfErr & Discrete and nominal & Indicator if player 1 has made an unforced error.  \\ \hline
P2UnfErr & Discrete and nominal & Indicator if player 2 has made an unforced error.  \\ \hline
P1NetPoint & Discrete and nominal & Indicator if player 1 has approached the net.  \\ \hline
P2NetPoint & Discrete and nominal & Indicator if player 2 has approached the net.  \\ \hline
P1NetPointWon & Discrete and nominal & Indicator if player 1 has won the point by approaching the net.  \\ \hline
P2NetPointWon & Discrete and nominal & Indicator if player 2 has won the point by approaching the net.  \\ \hline
P1BreakPoint & Discrete and nominal & Indicator if player 1 has a break point.  \\ \hline
P2BreakPoint & Discrete and nominal & Indicator if player 2 has a break point.  \\ \hline
P1BreakPointWon & Discrete and nominal & Indicator if player 1 wins a break point.  \\ \hline
P2BreakPointWon & Discrete and nominal & Indicator if player 2 wins a break point.  \\ \hline
Speed\_MPH & Continuous and ratio & Speed of the serve in miles per hour.    \\ \hline
P1BreakPointMissed & Discrete and nominal & Indicator if player 1 lose a break point.  \\ \hline
P2BreakPointMissed & Discrete and nominal & Indicator if player 2 lose a break point.  \\ \hline
Serveindicator & Discrete and Ordinal & Which player serves. Player 1 or player 2.     \\ \hline
ServeNumber & Discrete and Ordinal & Which serve the serving player succeed. First or second.    \\ \hline
WinnerType & Discrete and Ordinal & Indicator if the serving player won with the serve.    \\ \hline
WinnerShotType & Discrete and Ordinal & Indicator if the  player which won the point won with backhand or forehand.    \\ \hline
P1DistanceRun & Continuous and ratio & Length the player 1 covered in the point in meters.    \\ \hline
P2DistanceRun & Continuous and ratio & Length the player 2 covered in the point in meters.    \\ \hline
RallyCount & Discrete and ratio & Number of times the ball crosses the net in a point.    \\ \hline
ServeWidth & Discrete and Ordinal & Indicator at how wide the successful serve was.  The feature can take the values: B (Body), BC (Body/center), BW (Body/wide), C (Center) or W (Wide).   \\ \hline
ServeDepth & Discrete and Ordinal & Indicator at how deep the successful serve was. The feature can take the values: CTL (Close to line) or NCTL (Not close to line).    \\ \hline
ReturnDepth & Discrete and Ordinal & Indicator at how deep the successful return ball was. The feature can take the values: D (Deep) or ND (Not deep).   \\ \hline
\end{tabularx}
\end{table}


\section{Results}
This section presents the results of the Point Winner classification models analyzed and analyze the performance of the final models through feature importance and evaluation metrics.

\subsection{Point winner model first serve}
For the first serve Point winner model, there is a class imbalance of roughly 73.2 \% winning the first serve vs / 27.8 \% losing the first serv. The baseline is calculated by the number of times the server wins the point in percentage. Below in Table \ref{Point_Serve_1} are the average metric scores from 10-fold cross-validation applied to the described machine learning methods using standard parameters.

\begin{table}[h]
\begin{center}
\caption{Average values of 10-fold-cross-validation for the evaluation metrics for the first serve Point Winner models, on training and validation data.}
\begin{tabular}{|
>{\columncolor[HTML]{F2F2F2}}l |
>{\columncolor[HTML]{F2F2F2}}l |
>{\columncolor[HTML]{F2F2F2}}l |
>{\columncolor[HTML]{F2F2F2}}l |
>{\columncolor[HTML]{F2F2F2}}l |
>{\columncolor[HTML]{F2F2F2}}l |}
\hline
\textbf{Model}      & \textbf{Accuracy score {[}\%{]}} & \textbf{Recall {[}\%{]}} & \textbf{Precision  {[}\%{]}} & \textbf{F1-score {[}\%{]}} & \textbf{Roc-Auc {[}\%{]} } \\ \hline
\rowcolor[HTML]{F2F2F2}
Baseline            & 73.2                    & \multicolumn{1}{c|}{-} & \multicolumn{1}{c|}{-}   & \multicolumn{1}{c|}{-} & \multicolumn{1}{c|}{-} \\ \hline
XGBoost             & 73.2                             & 0.0               &  16.8              & 0.1             & 57.3            \\ \hline
Adaboost            &  73.2                                & 0.0               & 0.0                  & 0.0                 & 57.3            \\ \hline
Random Forrest      & 73.2                             & 0.0               & 10.0                & 0.0                 & 57.6            \\ \hline
Logistic Regression & 73.2                             & 0.0               & 31.0              & 0.2             & 57.6            \\ \hline
\end{tabular}
\label{Point_Serve_1}
\end{center}
\end{table}

From Table \ref{Point_Serve_1} it can be seen that the general model performance is almost identical. All the models scores roughly equally to the baseline in the accuracy score, however performs very poorly in F1, recall and precision score, which means the models rarely or never actually predicts the returner as the winner. Logistic Regression does however score higher in precision, however due to the low recall score the result is quite unreliable. XGBoost is chosen for further evaluation and hyperparameter tuning, due to the historical performance of XGBoost \cite{nielsen2016tree}, and its multiple hyper parameters which can be configured and where the optimal hyper parameters are found by random search.
Here it can be noted that the \textit{scale\_pos\_weight} is 1.3 which scales the gradient for the under represented class and encourages the loss function to focus on correcting mistakes of predicting for this class. Finally the tuned model is trained on the training and validation data and tested on the test data giving the following results:

\begin{table}[h]
\begin{center}
\caption{Tuned XGBoost for first serve PointWinner predictions on the test data}
\begin{tabular}{|
>{\columncolor[HTML]{F2f2F2}}l |
>{\columncolor[HTML]{F2f2F2}}l |
>{\columncolor[HTML]{F2f2F2}}l |
>{\columncolor[HTML]{F2f2F2}}l |
>{\columncolor[HTML]{F2f2F2}}l |
>{\columncolor[HTML]{F2f2F2}}l |}
\hline
\textbf{Model} & \textbf{Accuracy score {[}\%{]} } & \textbf{Recall {[}\%{]}} & \textbf{Precision {[}\%{]} } & \textbf{F1 score {[}\%{]} } & \textbf{Roc-Auc  {[}\%{]} } \\ \hline
\rowcolor[HTML]{F2F2F2}
Baseline            & 73.2                    & \multicolumn{1}{c|}{-} & \multicolumn{1}{c|}{-}   & \multicolumn{1}{c|}{-} & \multicolumn{1}{c|}{-} \\ \hline       
XGBoost        & 73.4                             & 1.2               & 43.7                  & 2.3                 & 56.9            \\ \hline
\end{tabular}
\label{OldXGBoostFS}
\end{center}
\end{table}

\begin{figure}[h]
    \centering
    \includegraphics[scale=0.8]{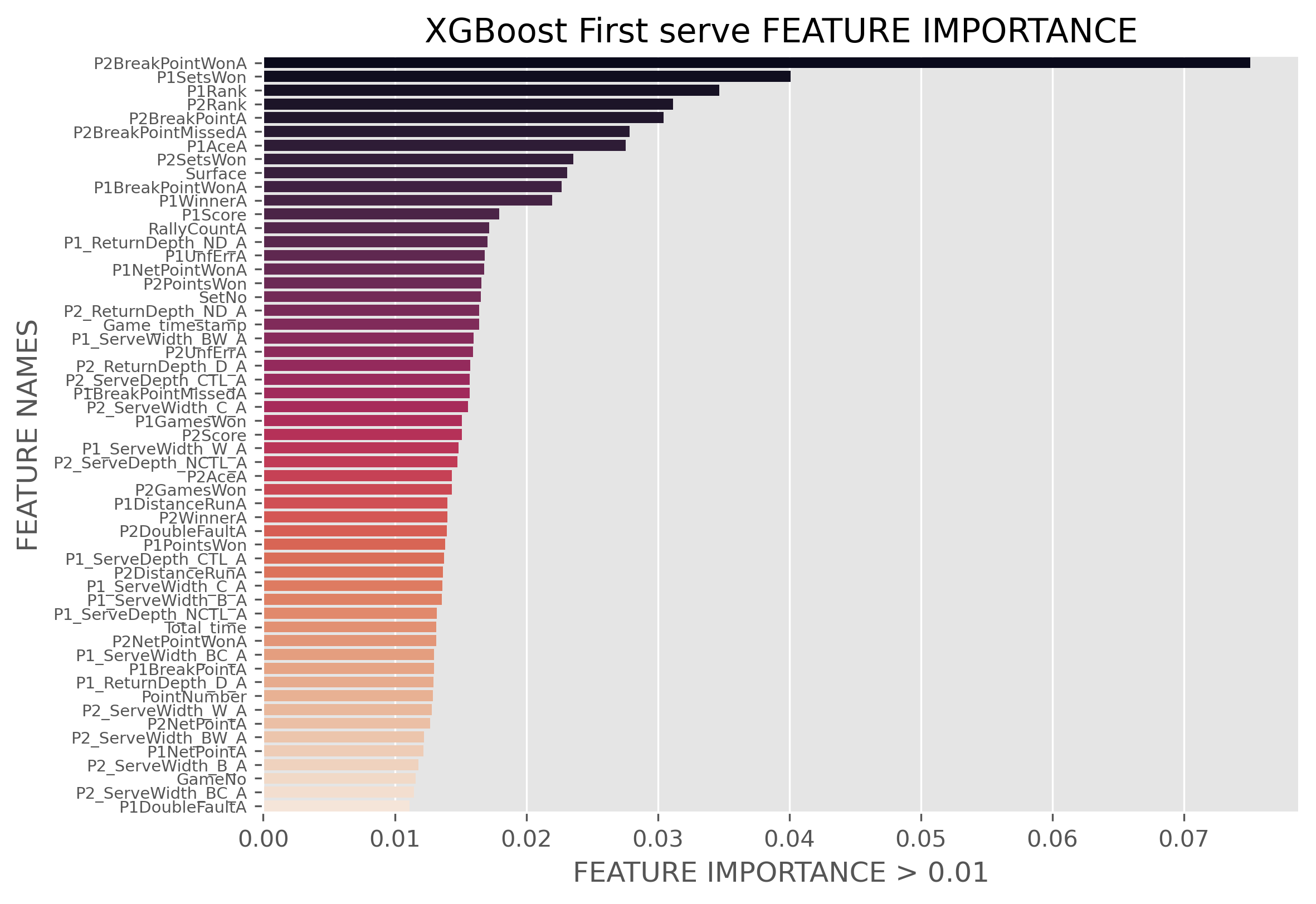}
    \caption{Feature importance (Gain) $>$ 0.01 for tuned XGBoost for first serve PointWinner model, sorted by value} 
    \label{fig:Serv1}
\end{figure}

 The model achieves a slight gain in accuracy score compared to the baseline, but in general the overall model performance is questionable in terms of the evaluation metrics. The model performs overall poorly at predicting when the returner wins however, the Roc-Auc still indicates that it is slightly better than a random prediction. In order to examine how the different features contributes to the model, the gain of each feature is presented in figure \ref{fig:Serv1}, which describes the relative contribution of the corresponding feature to the model calculated by taking each feature’s contribution for each tree in the model. It can be seen that numerous features has an impact on the model, resulting in the average individual contribution being low. A closer examination of the most contributing features shows that some of the more important features are \textit{P1SetsWon} and \textit{P2SetsWon} which describes the number of sets the server and returner has won at the point being predicted, and thereby their overall performance in the match. The another feature describing number of breakpoints won or missed for the returner given by \textit{P2BreakPointMissedA} and \textit{P2BreakPointWonA} is also relative important, which unlike the \textit{P2SetsWon}, describes how the returner has played in critical points of the match. The rank of each player given by the features \textit{P1Rank} and \textit{P2Rank} is also presented as important features, which indicates the rank of a player has impact on the outcome.

\subsubsection{Point Winner model for the second serve}
The second serve Point Winner model, is trained on all observations decided from the second serve. The class representation is roughly equal at 57\% P1 and 43\% P2, whereas the baseline is calculated by the number of times the server wins the point from the second serve in percentage. Below in Table \ref{2nd-serve-point-10-fold} the average metric scores from 10-fold cross-validation applied to the described machine learning methods using standard parameters can be seen.

\begin{table}[h]
\begin{center}
\caption{Second serve PointWinner models performance, metrics are averages from 10-fold cross-validation on training and validation data}
\label{2nd-serve-point-10-fold}
\begin{tabular}{|
>{\columncolor[HTML]{F2F2F2}}l |
>{\columncolor[HTML]{F2F2F2}}l |
>{\columncolor[HTML]{F2F2F2}}l |
>{\columncolor[HTML]{F2F2F2}}l |
>{\columncolor[HTML]{F2F2F2}}l |
>{\columncolor[HTML]{F2F2F2}}l |}
\hline
\textbf{Model}      & \textbf{Accuracy score {[}\%{]} } & \textbf{Recall {[}\%{]}  } & \textbf{Precision  {[}\%{]}   } & \textbf{F1-score {[}\%{]}} & \textbf{Roc-Auc  {[}\%{]}} \\ \hline
\rowcolor[HTML]{F2F2F2}
Baseline            & 57.2                    & \multicolumn{1}{c|}{-} & \multicolumn{1}{c|}{-}   & \multicolumn{1}{c|}{-} & \multicolumn{1}{c|}{-} \\ \hline
XGBoost             & 53.0                               & 44.8           & 52.3              & 48.3           & 54.1            \\ \hline
AdaBoost            & 52.8                             & 44.7           & 52.0               & 48.1             & 54.1            \\ \hline
Random Forest       & 53.0                               & 39.6           & 52.6              & 45.7             & 54.3            \\ \hline
Logistic Regression & 52.7                             & 39.5           & 52.3              & 45.2             & 53.8            \\ \hline
\end{tabular}
\end{center}

\end{table}

It shows that the model performances are almost identical. All models scores roughly 4\% lower in accuracy than the baseline. The recall, precision and F1 scores are significantly higher for the second serve than with the first serve, which indicates that it is easier to predict when the returner wins in the second serve compared to the first serve. However, the class distribution is more balanced for the second serve and an examination of Roc-Auc values shows that the models are slightly better than random guessing. The two best performing models are XGBoost and random forest, which performs almost equally in all metrics, but XGBoost is slightly better at predicting the returner. Therefore XGBoost is chosen as the best model for further evaluation and hyperparamter tuning.

Likewise for the Point Winner model for the first serve the hyperparamters are found by random search and the procedure can be seen in \cite{Github-respository}. The tuned XGBoost model is trained on the training and validation data and tested on the test data giving the following results:

\begin{table}[h]
\begin{center}
\caption{Tuned XGBoost for second serve Point Winner predictions on the test data}
\begin{tabular}{|
>{\columncolor[HTML]{F2f2F2}}l |
>{\columncolor[HTML]{F2f2F2}}l |
>{\columncolor[HTML]{F2f2F2}}l |
>{\columncolor[HTML]{F2f2F2}}l |
>{\columncolor[HTML]{F2f2F2}}l |
>{\columncolor[HTML]{F2f2F2}}l |}
\hline
\textbf{Model} & \textbf{Accuracy score {[}\%{]}} & \textbf{Recall {[}\%{]}} & \textbf{Precision {[}\%{]}} & \textbf{F1-score {[}\%{]}} & \textbf{Roc-Auc {[}\%{]}} \\ \hline
\rowcolor[HTML]{F2F2F2}
Baseline            & 57.2                    & \multicolumn{1}{c|}{-} & \multicolumn{1}{c|}{-}   & \multicolumn{1}{c|}{-} & \multicolumn{1}{c|}{-} \\ \hline
XGBoost        & 53.1                             & 44.7               & 50.6                  & 47.5                & 53.5            \\ \hline
\end{tabular}
\label{SecondS_test}
\end{center}
\end{table}

\begin{figure}[h]
    \centering
    \includegraphics[scale=0.7]{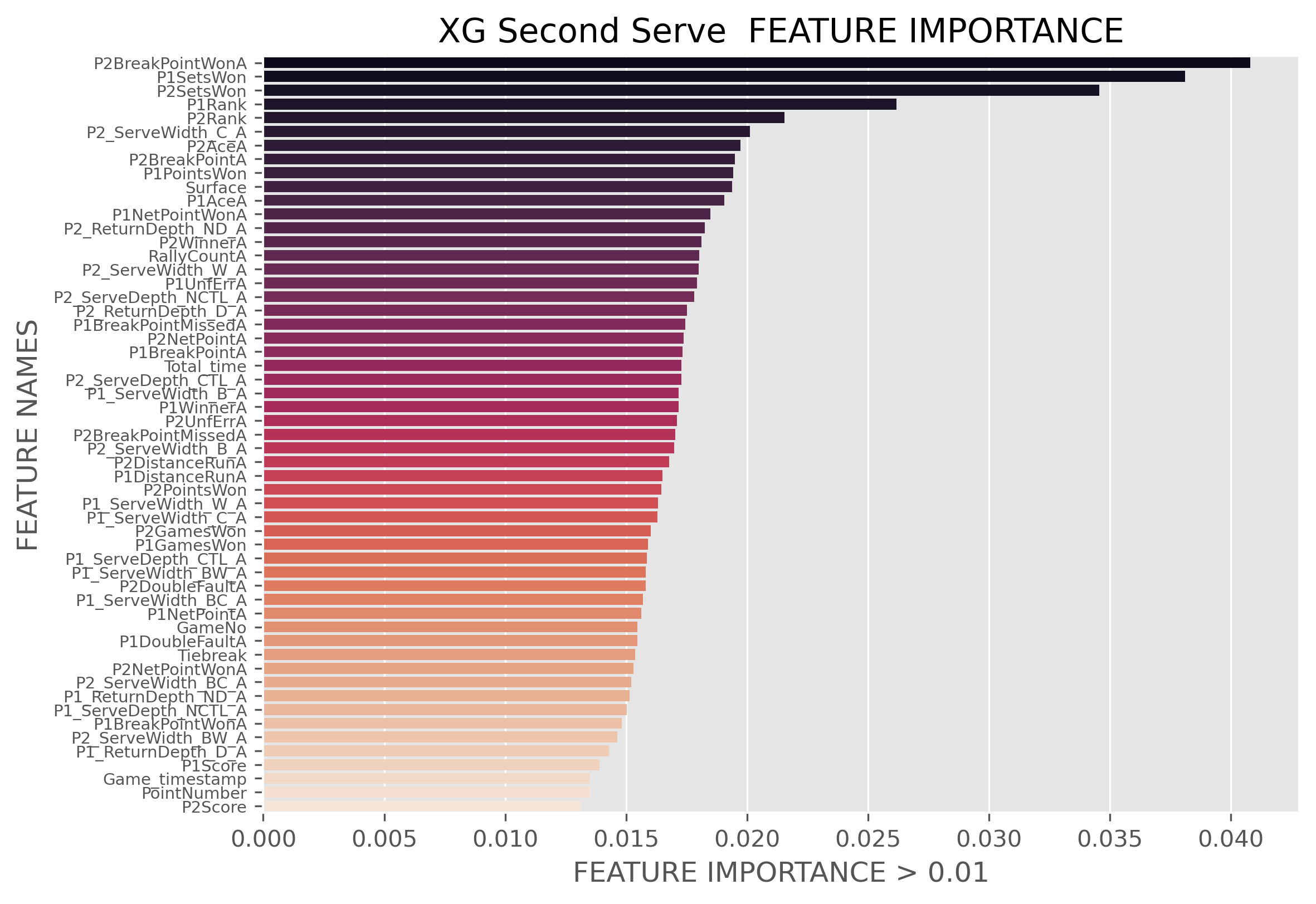}
    \caption{Feature importance (Gain) $>$ 0.01 for tuned XGBoost for the second serve point Winner model, sorted by value.} 
    \label{fig:Serv2}
\end{figure}

It can be seen that the tuned XGBoost model performs 5.1\% lower in accuracy score compared to the baseline. In general, the overall model performance is questionable in terms of the evaluation metrics. The Roc-Auc score indicates that the model is slightly better than a random guess. 
The model predicts 45\% of the points where the returner wins correctly out of all possible points where the returner wins. And of all points which the returner is predicted win, 50\% is correct. In order to examine which features which influence the model, the feature importance is shown in Figure \ref{fig:Serv2}.

Likewise the first serve model, as for the second serve model it can be seen that numerous features has an impact on the model, resulting in the average individual contribution being quite low. Some of the most important features based on gain are the accumulated features describing the score \textit{P2BreakPointWona}, \textit{P1SetsWon}, \textit{P2SetsWon} are strong indicators how well the given player has performed during the match. Match describing features like \textit{P1Rank} ,\textit{P2Rank} and \textit{Surface} also has a slightly higher impact on the model than the average feature. It can be seen that the numerous serve and return accumulated features contributes to the model, however none of them have a significant impact.

\section{Conclusion}
From the analysis, it is clear, that all the models have their own challenges. The Point Winner model performed equal or worse than their baselines, and the first serve Point Winner model had significant difficulties making predictions for the returner. The first serve Point Winner model primary challenge was that it did not predict the returner as the winner in any of the points. Therefore the accuracy score was almost the same as the baseline. When the important features were examined could it be seen that multiple attributes had influence on the model, however the general contribution was very low. This could indicate that none of the attributes actually having any influence on the model, and its predictions are almost solely based on the class. The second serve Point Winner model did a better at predictions for the returner, however the overall accuracy of the model was roughly 5\% lower than its baseline. The Roc-Auc was just above 0.5 which indicated that model was not performing much better than a random guess. An examination of the features showed, as the first serve Point Winner model, that average feature contribution was very low, which could indicate that none of the features actually have an impact on the models performance.

It was found the features which proved to influence the models the most was either score related or the rank of the players. The accumulated features for the serve and return position had no significant influence. It has therefore not been possible to identify factors which gives strategic advances in tennis. The approach presented can be used to predict the outcome of points, however the results are not ideal with the current data-set for examining which factors are important for the outcome of tennis matches.



\printbibliography[heading=bibintoc]
\end{document}